\documentclass[11pt]{llncs}
\usepackage[margin=1.1in]{geometry}

\usepackage[english]{babel}
\usepackage[utf8]{inputenc}
\usepackage{amsmath}
\usepackage{amsfonts}
\usepackage{graphicx}
\usepackage{rotating}
\usepackage[table]{xcolor} 
\usepackage{lipsum}
\usepackage{longtable}

\DeclareMathOperator*{\argmax}{arg\,max}
\newcommand{\aptastruct}{AptaTRACE}

\title{\aptastruct: Elucidating Sequence-Structure Binding Motifs by Uncovering Selection Trends in HT-SELEX Experiments}

\author{Phuong Dao\inst{1}\fnmsep\thanks {Equal contribution, these authors are listed in alphabetical order} \and Jan Hoinka\inst{1}\fnmsep$^\star$ \and Yijie Wang\inst{1} \and Mayumi Takahashi\inst{2} \and Jiehua Zhou\inst{2} \and Fabrizio Costa\inst{3} \and John Rossi\inst{2} \and John Burnett\inst{2} \and Rolf Backofen\inst{3} \and Teresa M.Przytycka\inst{1}}
\institute{National Center of Biotechnology Information, National Library of Medicine, NIH, Bethesda MD 20894, USA \email{przytyck@ncbi.nlm.nih.gov} \and
Department of Molecular and Cellular Biology, Beckman Research Institute of City of Hope, CA, USA \and
Bioinformatics Group, Department of Computer Science, University of Freiburg, Georges-Kohler-Allee 106, 79110 Freiburg, Germany
}

\date{\today}

\begin{document}
\maketitle
\begin{abstract}
Aptamers, short synthetic RNA/DNA molecules binding specific targets with high affinity and specificity, are utilized in an increasing spectrum of bio-medical applications. 
Aptamers  are identified \textit{in vitro} via the Systematic Evolution of Ligands by Exponential Enrichment (SELEX) protocol. SELEX selects binders through an iterative process that, starting from a  pool of random ssDNA/RNA sequences, amplifies target-affine species through a series of selection cycles. HT-SELEX, which combines SELEX with high throughput sequencing, has recently transformed aptamer development and has opened the field to even more applications. HT-SELEX is capable of generating over half a billion data points, challenging computational scientists with the task of identifying aptamer properties such as sequence structure motifs that determine binding. 
While currently available motif finding approaches suggest partial solutions to this question, none possess the generality or scalability required for HT-SELEX data, and they do not take advantage of important properties of the experimental procedure.\\\\
We present \aptastruct , a novel approach for the identification of sequence-structure binding motifs in HT-SELEX derived aptamers.  Our approach leverages the experimental design of the SELEX protocol and identifies sequence-structure motifs that show a signature of selection. Because of its unique approach, \aptastruct\ can   uncover motifs even when these are present in only a minuscule fraction of the pool. Due to these features, our method can help to reduce the number of selection cycles required to produce aptamers with the desired properties, thus reducing cost and time of this rather expensive procedure. The performance of the method on simulated and real data indicates that \aptastruct\ can detect sequence-structure motifs even in highly challenging data. 
\end{abstract}

\newpage
\setcounter{page}{1}

\section{Introduction}
Aptamers are short RNA/DNA molecules capable of binding, with high affinity and specificity, a specific target molecule via sequence and structure features that are complementary to the biochemical characteristics of the target's surface. The utilization of aptamers in a multitude of biotechnological and medical sciences has recently  dramatically increased. While only 80 aptamer related publications were added to Pubmed in the year 2000, this number has since roughly doubled every 5 years, with 207 records added in 2005 alone, 565 additional inclusions in 2010, and as many as 957 new manuscripts indexed in 2014. This astonishing trend is in part attributable to the considerable diversity of possible targets which span from small organic molecules \cite{Kim_Gu_2013}, over transcription factors \cite{Jolma2010} and other proteins or protein complexes \cite{Berezhnoy_2012}, to the surfaces of viruses \cite{Binning_Wang_Luthra_Shabman_Borek_Liu_Xu_Leung_Basler_Amarasinghe_2013} and entire cells \cite{Shi_Cui_He_Guo_Wang_Ye_Tang_2013}. This broad range of targets makes aptamers suitable candidates for a variety of applications ranging from molecular biosensors \cite{Zichel_Chearwae_Pandey_Golding_Sauna_2012}, to drug delivery systems \cite{duan2015}, and antibody replacement \cite{2004Degeneration} to just name a few.\\
\indent While the specifics vary depending on the  target, aptamers are typically identified through the Systematic Evolution of Ligands by Exponential Enrichment (SELEX) protocol \cite{Ellington_Szostak_1990}. SELEX leverages the well established paradigm of \textit{in vitro} selection by repetitively enriching a pool of initially random sequences (species) with those that strongly bind a target of interest. Specifically, based on the assumption that a large enough initial pool of randomized (oligo)nucleotides contains some species with favorable sequence and structure allowing for binding to the target, these binders are then selected for through a series of selection cycles. Each such cycle involves (a) incubating the pool with the target molecules, (b) partitioning target-bound species from non-binders and (c) removing the latter from the pool, followed by (d) elution of the bound fraction from the target, and (e) amplifying the remaining sequences via polymerase chain reaction (PCR) to form the input for the subsequent round. After a target-specific number of selection cycles, the final pool is then used to extract dominating, putatively high-affinity species, via traditional cloning experiments, computational analysis, and binding affinity assays. Depending on their intended application, favorable binders are often further post-processed \textit{in vitro} to meet additional requirements such as improved structural stability or reducing the size of the aptamer to the relevant binding region.\\
\indent Another reason for the resurgence of interest in aptamer research relates to the utilization of affordable next-generation sequencing technologies along with traditional SELEX. This novel protocol, called HT-SELEX, combines Systematic Evolution of Ligands by Exponential Enrichment with high-throughput sequencing. In HT-SLELEX, after certain (or all) rounds of selection (including the initial pool), aptamer pools are split into two samples, the first of which serves as the starting point for the next cycle whereas the latter is sequenced. The resulting sequencing data, consisting of 2-50 million sequences per round, is then analyzed \textit{in silico} in order to identify candidates that experience exponential enrichment throughout the selection \cite{hoinkaRECOMB,Fastaptamer2015}. \\
\indent The massive amount of sequencing data produced by HT-SELEX opens the opportunity for the study of many aspects of the protocol that were either not accessible in traditional SELEX or that could be realized more accurately given hundreds of millions of data points. One of the most challenging of these problems is the discovery of aptamer properties that facilitate binding to the target. However, development of universal methods for the analysis of HT-SELEX data is challenged by the vast diversity of selection conditions such as temperature, salt concentration and the number of targets in the solution to just name a few. Further, each of the stages (a-e) comprising one selection cycle can be accomplished by a variety of technologies. For instance, choosing between open PCR and droplet PCR for the amplification step has been shown to have a great impact on the diversity of the amplification product \cite{krylov2006,krylov2015,Burnett2015}. Even more importantly, the complexity of the target molecule is also of great relevance. As a case in point, it has been  shown that \textit{in vitro} selection against transcription factors, and other molecules that are evolutionary optimized to efficiently recognize specific DNA/RNA targets, requires only a small number of rounds in order to produce high quality aptamers \cite{Kupakuwana_Crill_McPike_Borer_2011,Jolma2010}. On the other side of the spectrum, in the case of CELL-SELEX, a variation of SELEX in which the pool is incubated with entire cells, the number of required selection cycles and the amount of non-specific binders that emerge during selection is significantly larger. Indeed, such a target can in general accommodate a multitude of binding sites, each exposing different binding preferences and leading to a parallel selection towards unrelated binding motifs. Current motif finding algorithms however, have not been designed with these challenges in mind and the need for the development of novel approaches that address the characteristics specific to the SELEX protocol has become highly relevant.\\
\indent \textbf{Traditionally, motif discovery} has been defined as the problem of finding a set of common sub-sequences that are statistically enriched in a given collection of DNA, RNA, or protein sequences. To date, a large variety of computational methods in this area has been published (see \cite{Tompa2005,Das2007,Zambelli2013} for a comprehensive review). In one of the earlier works, Lawrence and Reilly \cite{Lawrence1990} introduced an Expectation Maximization (EM) based algorithm for finding motifs from protein sequences. This approach has been consequently adopted by various other methods \cite{Sinha2004,Reid2011} including MEME \cite{Bailey1995} -- one of the most widely used programs in this category. Lawrence \textit{et al.} also introduced a Gibbs sampling approach for motif identification \cite{Lawrence1993} which laid the grounds for other methods such as AlignACE \cite{Roth1998}, MotifSampler \cite{Thijs2002}, and BioProspector \cite{Liu2001} based on this general technique.
In addition, numerous approaches have been designed based on efficient counting of all possible $k$-mers in a data set followed by a statistical analysis of their enrichment. Representatives for this category include Weeder \cite{Pavesi2004}, DREME \cite{Bailey2011}, YMF \cite{Sinha2002}, MDScan \cite{Liu2002}, and Amadeus \cite{Linhart2008}. Kuang \textit{et al.} designed a kernel based technique around a set of similar $k$-mers with a small number of mismatches to extract short motifs in protein sequences \cite{Kuang2012}.  
Another group of algorithms that also allows for elucidation of motifs with mismatches is built on suffix tree techniques  (Sagot \cite{Sagot98}, Pavesi \textit{et al.} \cite{Pavesi2001}, and Leibovich \cite{Leibovich2013}). Furthermore, regression based methods have been developed that take additional information, such as the affinity of the input sequences or the genomic regulatory contexts into account. These include, but are not limited to MatrixREDUCE \cite{Foat2006}, PREGO \cite{Tanay2006}, ChIPMunk \cite{Kulakovskiy2010}, and SeqGL \cite{Setty2015}. For more information, we refer the reader to Weirauch \textit{et al.} \cite{Weirauch2013} for a comprehensive evaluation of many of the above techniques. Finally, a number of approaches for the identification of sequence motif in HT-SELEX data targeting transcription factors (TF-SELEX) have been published. One representative of this category is BEEML \cite{Zhao2009}, which is, to our knowledge, the first computational method for finding motifs on this type of high-throughput sequencing data. Assuming the existence of a single binding motif, the method aims at fitting a binding energy model to the data which combines independent attributes from each position in the motif with higher order dependencies. Another method by Jolma \textit{et al}. approaches the problem by using $k$-mers to construct a position weight matrix in order to infer the binding models \cite{Jolma2010,Jolma2013}. Similarly, Orenstein \textit{et al.} \cite{Orenstein022277} also uses a $k$-mer approach based on frequencies from a single round of selection to identify binding motifs for transcription factor HT-SELEX data. Notably, despite of HT-SELEX' capability of generating data from multiple rounds of selection, all currently existent methods are based on the analysis of only a single selection cycle. However, choosing the round for optimal motif elucidation is not always trivial, and while some effort has been made to address this question (see for example Orenstein and Shamir \cite{Orenstein022277}, Jolma \textit{et al.} \cite{Jolma2010}) this decision is ultimately left to the user.\\
\indent \textbf{The search for motifs in the context of RNA} sequences faces another dimension in complexity as binding of ssDNA and RNA molecules is known to be sequence and structure dependent. In particular, it has been proposed that binding regions in those molecules tend to be predominantly single stranded \cite{Schudoma2010}. MEMERIS \cite{Hiller2006}, for instance, leverages this assumption by weighting nucleotides according to their likelihood of being unpaired. These positional weights then guide MEME to focus the motif search on loop regions. In contrast, RNAcontext \cite{Morris2010} divides the single stranded contexts into known secondary substructures such as hairpins, bulge loops, inner loops, and stems. Consequently, RNAcontext is capable of reporting the relative preference of the structural context along with the primary structure of the potential motif. Recently, Hoinka et al. introduced AptaMotif \cite{Hoinka2012} a method to discover sequence–structure motifs from SELEX derived aptamers. This method utilizes information about the structural ensemble of aptamers obtained by enumerating of all possible structures within a user-defined energy range from the Minimum Free Energy (MFE) structure. By representing each aptamer by the set of its unique substructures (i.e. hairpins, bulge-loops, inner-loops, and multibranch loops), AptaMotif applies an iterative sampling approach combined with sequence-structure alignment techniques to identify high-scoring seeds which are consequently extended to motifs over the full data set. However, AptaMotif was designed for sequencing data obtained from traditional SELEX, under the assumption that this data predominantly consists of motif containing sequences. Subsequently, APTANI \cite{APTANI} extended AptaMotif to handle larger sequence collections via a set of parameter optimizations and sampling techniques, but it also expects a high ratio of motif occurrences. \\
\indent \textbf{Still, none of the above mentioned methods} address the full spectrum of challenges when analyzing data from HT-SELEX selections. First, none of these approaches, as currently implemented, scales well with the data sizes produced by modern high throughput sequencing experiments. Next, only a few of the methods consider the existence of secondary motifs while the majority operates under the assumption that only a singe primary motif is present in the data. This assumption might apply to TF-SELEX, but it cannot be generalized to common purpose HT-SELEX but where one should  consider many motifs of possibly similar strength. Furthermore, secondary structure information, which has proven effective in guiding the motif search to biologically relevant binding sites, is not included in most of these methods. A notable exception is RNAContext which can handle relatively large data sets but suffers from the single motif assumption that cannot be easily removed. Finally, none of these approaches attempt to utilize the full scope of the information produced by modern HT-SELEX experiments that includes sequencing data from multiple rounds of selection.\\
\indent \textbf{In order to close this gap}, we have developed \aptastruct, a method for the identification of sequence-structure motifs for HT-SELEX that utilizes the available data from all sequenced selection rounds, and which is robust enough to be applicable to a broad spectrum of RNA/ssDNA HT-SELEX experiments, independent of the target's properties. Furthermore, \aptastruct \ is not limited to the detection of a single motif but capable of elucidating an arbitrary number of binding sites along with their corresponding structural preferences. \aptastruct\ approaches the sequence-structure motif finding problem in a novel and unique way. Unlike previous methods, it does not rely on aptamer frequency or its derivative - cycle-to-cycle enrichment. Aptamer frequency has been recently shown to be a poor predictor of aptamer affinity \cite{hoinkaRECOMB,soh2010,Thiel2012}, and while cycle-to-cycle enrichment has shown a somewhat better performance, the choice of the cycles to compare is not obvious and does not always allow for extraction of sequence-structure motifs. In contrast, our method builds on tracing the dynamics of the SELEX process itself to uncover motif-induced selection trends.\\
\indent We applied \aptastruct\ to sequencing data obtained from realistically simulating SELEX over 10 rounds of selection (4 million sequences per round) with known binding motifs as well as to an \textit{in vitro} cell-SELEX experiment over 9 selection cycles (40 million sequences per cycle). In both cases, our method was successful in extracting highly significant sequence-structure motifs while scaling well with the 10-fold increase in data size.
\section{Results} 
We start with a high-level outline of the method, followed by a more detailed description. Next, we use simulated data produced with a novel, extended version of our AptaSim program \cite{hoinkaNAR} to compare the performance of \aptastruct\ to other methods that can handle similar data sizes or incorporate secondary structure into their models. Finally, we show our results of applying \aptastruct\ to an \textit{in vitro} selection consisting of high-throughput data from 9 rounds of cell-SELEX \cite{Burnett2015}. 
\subsection{Top Level Description of the Algorithm}
Our method builds on accepted assumptions regarding the general HT-SELEX procedure. First, we assume that the affinity and specificity of aptamers are mainly attributed to a combination of localized sequence and structural features that exhibit complementary biochemical properties to a target's binding site. Given a large number of molecules in the initial pool it is expected that such binding motifs are embedded in multiple, distinct aptamers. Consequently, during the selection process, aptamers containing these highly target-affine sequence-structure motifs will become enriched as compared to target non-specific sequences. Notably, under these assumptions, aptamers that contain only the sequence motif without the appropriate structural context are either not enriched at all or enriched to a much lower degree. The second critical assumption we make is the existence of a multitude of sequence-structure binding motifs that either compete for the same binding site, or are binding to different surface regions of the target \cite{Zichel_Chearwae_Pandey_Golding_Sauna_2012}. \\
\indent Leveraging the above properties of the SELEX protocol, \aptastruct\ detects sequence-structure motifs by identifying sequence motifs which undergo selection towards a particular secondary structure context. Specifically, we expect that in the initial pool the structural contexts of each $k$-mer are distributed according to a background distribution that can be determined from the data. However, for sequence motifs involved in binding, in later selection cycles, this distribution becomes biased towards the structural context favored by the binding interaction with the target site. Consequently, \aptastruct \ aims at identifying sequence motifs whose tendency of residing in a hairpin, bugle loop, inner loop, multiple loop, danging end, or of being paired converges to a specific structural context throughout the selection. To achieve this, for each sequenced pool we compute the distribution of the structural contexts of all possible $k$-mers (all possible nucleotides  sequences of length $k$) in all aptamers. \\
\indent Next, we use the relative entropy (KL-divergence) to estimate, for every $k$-mer, the change in the distribution of its secondary structure contexts ($K$-context distribution, for short) between any cycle to a later cycle. The sum of these KL-divergence scores over all pairs of selection cycles defines the context shifting score for a given $k$-mer. The context shifting score is thus an estimate of the selection towards the preferred structure(s). Complementing the context shifting score is the $K$-context trace, which summarizes the dynamics of the changes in the $K$-context distribution over consecutive selection cycles. \\
\indent In order to assess the statistical significance of these context shifting scores, we additionally compute a null distribution consisting of context shifting scores derived from $k$-mers of all low-affinity aptamers in the selection. This background is used to determine a $p$-value for the structural shift for each $k$-mer. Predicted motifs are then constructed by aggregating overlapping $k$-mers under the restriction that the structural preferences in the overlapped region are consistent. Finally, Position Specific Weight Matrices (PWM), specifically their sequence logos representing these motifs, along with their motif context traces (the average $k$-context traces of the $k$-mers used in the PWM construction) are reported to the user.
\begin{sidewaysfigure} 
	\includegraphics[width=\textwidth]{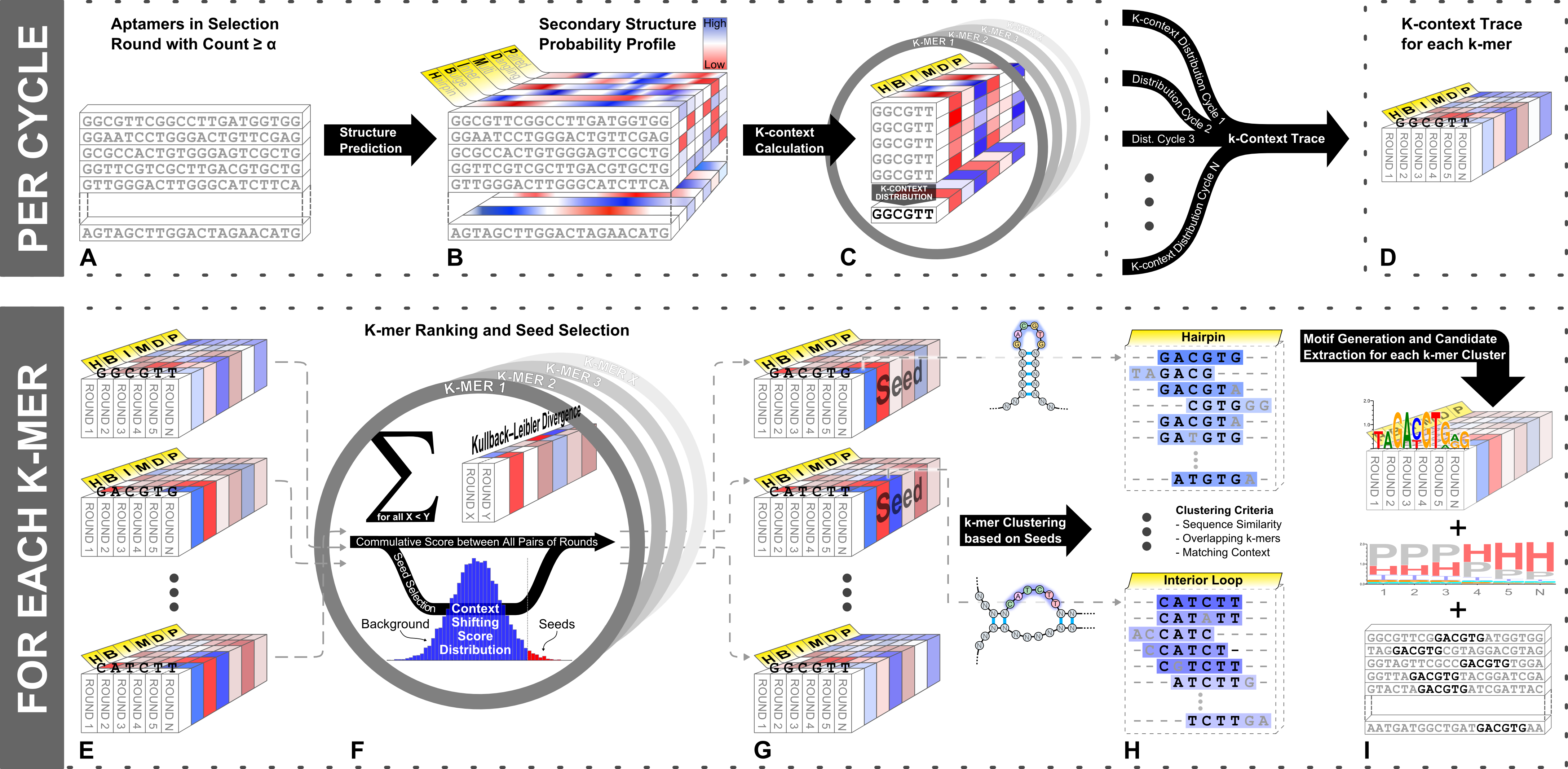}
    \caption{Schematic overview of our \aptastruct\ method. \textbf{(A)} For each cycle, all sequences with frequency above a user defined threshold $\alpha$ are selected as input. \textbf{(B)} Computation of secondary structure probability profiles for each aptamer using SFOLD. For each nucleotide the profile describes the probability of residing in a hairpin, bugle loop, inner loop, multiple loop, danging end, or of being paired. \textbf{(C)} $K$-context and $K$-context distribution calculation for each $k$-mer. \textbf{(D)} Generation of the $K$-context trace for each $k$-mer. \textbf{(E-G)} $K$-mer ranking and statistical significance estimation. Given any two selection cycles the relative entropy (KL-divergence) is used to estimate the change in the distribution of its $K$-context distribution. The sum of these KL-divergence scores over all pairs of selection cycles defines the context shifting score for a given $k$-mer. In order to assess the statistical significance of these context shifting scores, a null distribution is computed consisting of context shifting scores derived from $k$-mers of all low-affinity aptamers in the selection (frequency $\leq \alpha$). This background is used to determine a $p$-value for the structural shift for each $k$-mer. Top scoring $k$-mers are selected as seeds. (H) Predicted motifs are constructed by aggregating $k$-mers overlapping with the seed under the restriction that the structural preferences in the overlapped region are consistent. \textbf{(I)} Position Specific Weight Matrices representing these motifs, along with their $K$-context traces are reported to the user.}
	\label{fig:overview}
\end{sidewaysfigure}
\subsection{Detailed Description of \aptastruct \ }
\aptastruct \ takes as input the sequencing results from all, or a subset of selection cycles from an HT-SELEX experiment and outputs a list of position specific weight matrices (PWMs) along with a visual representation of the motifs structural context shift throughout the selection. \\\\
\textbf{$K$-context and $K$-context Distribution.} Any individual occurrence of a $k$-mer in an aptamer has a specific secondary structure context called $K$-context that depends on the structure of that particular aptamer. In what follows, let $K_i$ be the $i$-th $k$-mer (using an arbitrary indexing of all $4^k$ possible $k$-mers over the alphabet $\Omega_s=\{A,C,G,T\}$). In addition, let $R^x$ be the set of unique aptamers sequenced in selection round $x$ that have a frequency above a threshold $\alpha$ (this facilitates noise reduction - see computation of $p$-value below and  Fig. \ref{fig:overview}, A).\\
\indent First, for every aptamer $a$ of fixed length $n$ in $R^x$, we use SFold \cite{ding2001} to estimate the probability for each nucleotide in $a$ of being part of a hairpin (H), an inner loop (I), a bulge loop (B), a multi-loop (M), a dangling end (D), or being paired (P) (Fig. \ref{fig:overview}, B). Each aptamer $a$ is hence associated with a matrix of dimension $|\Omega_C| \times n$, where $\Omega_C=\{H,I,B,M,D,P\}$, in which rows correspond to a particular context $C$ while each column contains the context probabilities of the corresponding nucleotide in $a$. Next, we define the $K$-context of a $k$-mer occurrence in aptamer $a$ as the row-wise mean of the context probabilities over the matrix columns corresponding to the location of that $k$-mer in the aptamer sequence.\\
\indent Recall, that the main idea behind \aptastruct\ is to track the changes in secondary structure preferences of $k$-mers over the selection cycles. 
Capturing these secondary structure preferences should therefor take the entirety of $K$-contexts from all occurrences of a $k$-mer in a particular selection cycle into account. 
Thus, we define the $K$-context distribution of a $k$-mer $K_i$ in round $x$ as the averaged secondary structure profile of all $K$-contexts of $K_i$ over all aptamers in $R^x$. Formally, let $\mathbb{P}_i^x(C)$, where $C \in \Omega_C$, be the average probability of the structural context $C$ over all occurrences of the $k$-mer $K_i$ in all aptamers that meet the threshold criteria in round $x$. Then, the $K$-context distribution of $K_i$ in round $x$ is the vector $\mathbb{P}_i^x = [\mathbb{P}_i^x(H),\mathbb{P}_i^x(B),\mathbb{P}_i^x(I),\mathbb{P}_i^x(M),\mathbb{P}_i^x(D),\mathbb{P}_i^x(P)]$, normalized such that all entries sum up to one. (Fig. \ref{fig:overview} C). \\\\
\textbf{Analysis of the Shift of $K$-context Distributions during Selection.} If a $k$-mer forms part of a sequence-structure binding motif, its $K$-context distribution is expected to shift towards the context $C$ that is preferred for the binding interaction throughout the selection. In contrast, if a $k$-mer is not affected by selection, we expect little to no change in its context distribution over consecutive rounds. We  can capture this dynamics for any $k$-mer $K_i$ by its so called $K$-context trace $\mathbb{K}_i = [\mathbb{P}_i^0, \mathbb{P}_i^1,\ldots, \mathbb{P}_i^m \ ]$, defined as a vector tracking the $K$-context distribution over all $m$ selection cycles (Fig \ref{fig:overview} D). Our method consequently quantifies such shifts in the $K$-context distribution using the Kullback–Leibler divergence (relative entropy) -- a measure for the difference between two probability distributions. Here, for any $k$-mer $K_i$ the first distribution corresponds to the $K$-context distribution  $\mathbb{P}_i^x$ of an earlier round $x$ and the second to the $K$-context distribution $\mathbb{P}_i^y$ of a later selection cycle $y$. \\
\indent The KL-divergence between two appropriately chosen selection cycles might suffice, at least for some scenarios such as TF-SELEX, to capture the  shifts in $K$-context distributions. In practice however, for larger and more complex targets the selection landscape tends to be more complicated with various aptamers achieving peak enrichment at different selection cycles.
Thus makes it rather difficult to confidently chose such two presumably most informative cycles while ignoring remaining information. 
Therefore we  compute the cumulative KL-divergence between all pairs of sequenced pools. 
 In summary, we define the context shifting score $score(k_i)$ for $k$-mer $K_i$ as 
\begin{eqnarray*}
score(K_i)&=&\sum_{x=1}^{m-1}{\sum_{y=x+1}^{m}{D_{KL}(\mathbb{P}_i^y \| \mathbb{P}_i^x)}}\\
&=&\sum_{x=1}^{m-1}{\sum_{y=x+1}^{m}{\left[\sum_{C \in \Omega_C}{\mathbb{P}_i^y(C) \times \log{\frac{\mathbb{P}_i^y(C)}{\mathbb{P}_i^x(C)}}} \right]}}\\
\end{eqnarray*}
where  $D_{KL}(P \| Q)$ is the Kullback–Leibler divergence between two discrete distributions $P$ and $Q$.  To ensure statistical accuracy, the context shifting scores are only calculated for all $k$-mers with a count of at least $\beta$ individual occurrences in each pool (here, $\beta = 100$).\\\\
\textbf{Significance Estimation and $p$-value Computation.} While the context shifting score establishes a ranking of the $k$-mers in order of their overall change in secondary structure context, it does not provide any information over the statistical significance of that shift, i.e. it cannot distinguish between changes in response to the true selection pressure and changes associated with background noise such as non-binding species. These background species however are expected to occur in very low numbers throughout the selection. We leverage this property by using the context shifting scores of the $k$-mers from these low-count aptamers to construct a null distribution that is used to identify the significant context shifting scores for the full data set. In detail, we include all $k$-mer occurrences from aptamers that are not included in the previous generation of the context profiles, i.e. all aptamers below or equal to the user defined threshold $\alpha$. We note that the resulting null follows a log-normal distribution in our \textit{in vitro} experiment as well as for the simulation data presented in this study (see Section \ref{sec:experimental}).\\
\indent The above described procedure hence allows for the computation of a $p$-value for each $K$-context trace and we only retain those $K$-context traces with $p$-value 
below a user specifiable threshold (the default value is 0.01)
(Fig.\ref{fig:overview} E-G).\\\\
\textbf{Elucidating Sequence-Structure Motifs and Sequence Logos.} In the last step, \aptastruct\ proceeds to extract the final motifs by clustering similar and overlapping $k$-mers with correlating, statistically significant structural shifts together  (Fig \ref{fig:overview} H). This allows to uncover sequence-structure motifs that might extend over the chosen $k$-mer size and to build PWMs that summarize the motifs. 
Motif construction is accomplished iteratively. Until all $k$-mers have been assigned to a cluster, the most significant $k$-mer is first selected as seed, and all similar and highly overlapping $k$-mers with a $p$-value below the defined threshold and with comparable structural context are aggregated to the cluster.
The details of this relatively straightforward procedure are described in the Supplemenary Materials and Methods section \ref{sec:motifgeneration}. 
In a last step, the resulting motifs are reported to the user via their PWMs, sequence logos and their motif context traces, defined as the averaged $K$-context traces of those $k$-mers constituting the PWM. The set of aptamer candidates that satisfy both, the primary and secondary structure properties of the motifs, sorted by their statistical significance or frequency of occurrence is also included in the output (Fig. \ref{fig:overview} I).
\subsection{Results on Simulated Data}\label{sec:simulated}
{
\newcommand{\motifscalex}{0.7}
\newcommand{\motifscaley}{0.30}
\newcommand{\tracescalex}{0.7}
\newcommand{\tracescaley}{0.30}
\newcommand{\rowspace}{4ex}
\rowcolors{2}{blue!7}{white}
\begin{table}[t!]
	\caption{Sequence-structure motifs identified by \aptastruct\ from virtual SELEX given all 10 selection cycles including the initial pool as input. \aptastruct\ was able to recover all 5 motifs. Shown here are the identified sequence logos, the $k$-mer that scored highest in significance used for construction of each motif (seed) and its p-value, the abundance of the motif in the final selection round (Frequency), the first cycle at which the motif was detected ($C^*$), as well as the motif context trace throughout the selection from the initial pool to round 10.}
\begin{tabular}{m{3.5cm}m{2.2cm}m{2.5cm}m{2cm}m{0.8cm}m{4.5cm}m{0pt}}
\bf Sequence Logos &\bf Logo Seed &\bf Seed p-Value &\bf Frequency &\bf $C^*$ &\bf Motif Context Trace \\[1pt] \hline 
& & & & & \raisebox{-.5\height}{\scalebox{\tracescalex}[1]{\includegraphics{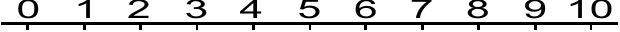}}} & \\
\raisebox{-.5\height}{\scalebox{\motifscalex}[\motifscaley]{\includegraphics[page=160]{logos.pdf}}}
& AGTCGG & 4.37E-36 & 24\% & 5 &
\raisebox{-.5\height}{\scalebox{\tracescalex}[\tracescaley]{\includegraphics[page=159]{logos.pdf}}} & 
\\[\rowspace]

\raisebox{-.5\height}{\scalebox{\motifscalex}[\motifscaley]{\includegraphics[page=182]{logos.pdf}}}
& GTGTAC & 1.91E-32 & 17.3\% & 5 &
\raisebox{-.5\height}{\scalebox{\tracescalex}[\tracescaley]{\includegraphics[page=181]{logos.pdf}}} &
\\[\rowspace]

\raisebox{-.5\height}{\scalebox{\motifscalex}[\motifscaley]{\includegraphics[page=224]{logos.pdf}}}
& TGGTTA & 2.19E-23 & 4.6\% & 9 &
\raisebox{-.5\height}{\scalebox{\tracescalex}[\tracescaley]{\includegraphics[page=223]{logos.pdf}}} &
\\[\rowspace]

\raisebox{-.5\height}{\scalebox{\motifscalex}[\motifscaley]{\includegraphics[page=144]{logos.pdf}}}
& GGAGCG & 2.83E-25 & 3.6\% & 8 &
\raisebox{-.5\height}{\scalebox{\tracescalex}[\tracescaley]{\includegraphics[page=143]{logos.pdf}}} &
\\[\rowspace]

\raisebox{-.5\height}{\scalebox{\motifscalex}[\motifscaley]{\includegraphics[page=188]{logos.pdf}}}
& GGAACT & 1.12E-27 & 1.9\% & 10 &
\raisebox{-.5\height}{\scalebox{\tracescalex}[\tracescaley]{\includegraphics[page=187]{logos.pdf}}} &
\\[\rowspace]
\end{tabular}
	\label{tab:simulationmotifs}
\end{table}
}
To test our new approach, we applied  \aptastruct\ to a data set generated by means of {\em in-silico }  SELEX  as no benchmarking set that could be used as a gold standard is currently available. To this end, we used an extension to our AptaSim program \cite{hoinkaNAR} designed to realistically simulate target-specific selection including, among other factors, species affinity, polymerase amplification and polymerase errors, and the effects of sampling from the selection pools for sequencing. Our current extension additionally allows for implanting sequence-structure motifs with well defined properties. We generated a data set of 4 million sequences per round containing 5 motifs (denoted here as motifs (a)-(e)), 5-8 nucleotides in length located predominantly in unpaired regions. 
Note that the motif sequence also occurs randomly in the background sequences, albeit in arbitrary structural contexts, and is hence not over-represented in the initial pool.Each motif was initially present in 100 different target-affine aptamer species and consequently selected for over 10 rounds of SELEX. A complete description of the simulation as well as the parameters used during \textit{in silico} SELEX are available in Supplementary Material and Methods Sections \ref{sec:simulation} and \ref{sec:parameters}, respectively.\\
\indent We applied \aptastruct, as well as DREME and RNAcontext to the data set to compare their capability of extracting these motifs. Since DREME and RNAcontext can only be applied to one selection round at a time, we provided these two approaches with data from the last selection cycle alone, choosing the initial pool as background when required. \aptastruct\ was applied to both, the reduced data set, as well as to all selection cycles. Notably, neither DREME nor RNAcontext are capable of handling 4 million sequences in a reasonable time frame, prompting us to sample 10\% of aptamers from the last and the unselected round as the input for DREME, and the 10000 most frequent and least frequent sequences of the last selection cycle for RNAcontext. The full scope of parameters used for these methods during the comparison are detailed in Supplementary Material and Methods \ref{sec:parameters}.\\
\indent Since RNAcontext's model assumes a single motif in the data, a direct comparison would not be fair for that software. Nonetheless, we examined the possibility of the method of identifying at least one binding site due to the large abundance of implanted motif (a) in the final selection round, however without success. Tab. \ref{tab:simulationcomparison} summarizes the results of \aptastruct\ when applied to the full dataset, as well as the to last selection cycle only, compared to DREMEs performance. While DREME failed to identify the low-affinity motif (e) as well as the shorter but more target-affine motif (c), \aptastruct\ was able to recover all motifs in both test scenarios.\\
\indent A more detailed summary of the sequence logos extracted by our approach on the full data set, including their motif context traces and statistical significance, is available in Tab. \ref{tab:simulationmotifs}. Interestingly, a visual inspection of the motif context trace (last column, Tab. \ref{tab:simulationmotifs}) points to the possibility of capturing most of these motifs at earlier cycles. Indeed, computing the selection round in which a motif was first detected by \aptastruct\ (column $C^*$, Tab.\ref{tab:simulationmotifs}), confirmed this expectation.
{
\newcommand{\motifscalex}{0.65}
\newcommand{\motifscaley}{0.30}
\newcommand{\rowspace}{4ex}
\rowcolors{2}{blue!7}{white}
\begin{table}
\caption{Comparison of \aptastruct\ against other methods based on simulated data. \aptastruct\ was applied to the entire dataset as well as to the last selection cycle only. While our method successfully identified all implanted motifs, DREME was only able of extracting 3 out of 5. We show the implanted motifs, their binding affinity used throughout the selection (B.A.) in the first two columns. The output PWMs produced by the tested methods that correspond to the implanted motifs are displayed in the remaining columns.}
\begin{tabular}{m{1cm}m{2.5cm}m{1.5cm}m{3.6cm}m{3.6cm}m{3cm}m{0pt}}
\bf & \bf Motif & \bf B.A. &\bf \aptastruct \ \ \ \ \ \ \ \ \ \ \ \ \ (all cycles) &\bf \aptastruct \ \ \ \ \ \ \ \ \ \ \ (last cycle)&\bf DREME \ \ \ \ \ \ \ \ \ \ \ (last cycle)& \\[1pt] \hline
(a) & GTGTACTA& 75\% &
\raisebox{-.5\height}{\scalebox{\motifscalex}[\motifscaley]{\includegraphics[page=160]{logos.pdf}}} &
\raisebox{-.5\height}{\scalebox{\motifscalex}[\motifscaley]{\includegraphics[page=250]{logos.pdf}}} &
\raisebox{-.5\height}{\scalebox{\motifscalex}[\motifscaley]{\includegraphics[page=433]{logos.pdf}}} &
\\[\rowspace]
(b) & AGTCGG& 70\% &
\raisebox{-.5\height}{\scalebox{\motifscalex}[\motifscaley]{\includegraphics[page=182]{logos.pdf}}} &
\raisebox{-.5\height}{\scalebox{\motifscalex}[\motifscaley]{\includegraphics[page=273]{logos.pdf}}} &
\raisebox{-.5\height}{\scalebox{\motifscalex}[\motifscaley]{\includegraphics[page=427]{logos.pdf}}} &
\\[\rowspace]
(c) & GTTAA& 60\% &
\raisebox{-.5\height}{\scalebox{\motifscalex}[\motifscaley]{\includegraphics[page=202]{logos.pdf}}} &
\raisebox{-.5\height}{\scalebox{\motifscalex}[\motifscaley]{\includegraphics[page=278]{logos.pdf}}} &
&
\\[\rowspace]
(d) & GGAACTC& 55\% &
\raisebox{-.5\height}{\scalebox{\motifscalex}[\motifscaley]{\includegraphics[page=212]{logos.pdf}}} &
\raisebox{-.5\height}{\scalebox{\motifscalex}[\motifscaley]{\includegraphics[page=287]{logos.pdf}}} &
\raisebox{-.5\height}{\scalebox{\motifscalex}[\motifscaley]{\includegraphics[page=431]{logos.pdf}}} &
\\[\rowspace]
(e) & GGAGCG& 50\% &
\raisebox{-.5\height}{\scalebox{\motifscalex}[\motifscaley]{\includegraphics[page=144]{logos.pdf}}} &
\raisebox{-.5\height}{\scalebox{\motifscalex}[\motifscaley]{\includegraphics[page=231]{logos.pdf}}} &
&
\end{tabular}
\label{tab:simulationcomparison}
\end{table}
\vspace*{-1cm}
}
\subsection{Results on Cell-SELEX Data}\label{sec:experimental}
Next, we applied \aptastruct\ to the results of an \textit{in-vitro} HT-SELEX experiment where the initial pool as well as 7 of 9 selection rounds have been sequenced, averaging 40 million aptamers per cycle (see  Section \ref{sec:cellselex} for a detailed description of the experimental procedure). We did not challenge DREME with this task, since this data set is 10-fold larger in size compared to the simulated selection, and even in the latter case DREME managed to only handle 10\% of the data. \aptastruct\ was able to successfully extract a total of 25  motifs, the five most frequent of which are shown in Tab. \ref{tab:openpcrmotifs}, and a full list is given in Supplementary Tab. \ref{tab:fullopenpcrmotifs}.\\
{
\newcommand{\motifscalex}{0.80}
\newcommand{\motifscaley}{0.30}
\newcommand{\tracescalex}{0.90}
\newcommand{\tracescaley}{0.30}
\newcommand{\rowspace}{4ex}
\rowcolors{2}{blue!7}{white}
\begin{table}
\caption{Five most frequent sequence-structure motifs as produced by \aptastruct\ on CELL-SELEX data. The sequence logo as well as the most frequent $k$-mer constituting the logo (Logo Seed) and its p-value are depicted for each motif. The motif context trace for the sequenced cycles (0,1,3,5,6,7,8,9) is shown in the last column.}
\begin{tabular}{m{3.5cm}m{2.5cm}m{3cm}m{2.2cm}m{4.1cm}m{0pt}}
\bf Sequence Logo &\bf Logo Seed &\bf Seed p-Value &\bf Frequency &\bf Motif Context Trace \\[2pt] \hline
& & & & \raisebox{-.5\height}{\scalebox{\tracescalex}[1]{\includegraphics{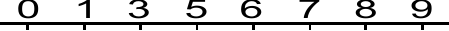}}} & \\
\raisebox{-.5\height}{\scalebox{\motifscalex}[\motifscaley]{\includegraphics[page=110]{logos.pdf}}}
& CTGTG & 5.31E-03 & 32.98 \% & 
\raisebox{-.5\height}{\scalebox{\tracescalex}[\tracescaley]{\includegraphics[page=109]{logos.pdf}}} & 
\\ [\rowspace]
\raisebox{-.5\height}{\scalebox{\motifscalex}[\motifscaley]{\includegraphics[page=124]{logos.pdf}}}
& TTATT & 7.16E-03 & 25.14 \% & 
\raisebox{-.5\height}{\scalebox{\tracescalex}[\tracescaley]{\includegraphics[page=123]{logos.pdf}}} & 
\\ [\rowspace]
\raisebox{-.5\height}{\scalebox{\motifscalex}[\motifscaley]{\includegraphics[page=126]{logos.pdf}}}
& GTTTA & 9.71E-03 & 16.77 \% & 
\raisebox{-.5\height}{\scalebox{\tracescalex}[\tracescaley]{\includegraphics[page=125]{logos.pdf}}} & 
\\ [\rowspace]
\raisebox{-.5\height}{\scalebox{\motifscalex}[\motifscaley]{\includegraphics[page=128]{logos.pdf}}}
& ATGTT & 3.66E-03 & 15.27 \% & 
\raisebox{-.5\height}{\scalebox{\tracescalex}[\tracescaley]{\includegraphics[page=127]{logos.pdf}}} & 
\\ [\rowspace]
\raisebox{-.5\height}{\scalebox{\motifscalex}[\motifscaley]{\includegraphics[page=130]{logos.pdf}}}
& GTGTC & 6.61E-03 & 13.95 \% & 
\raisebox{-.5\height}{\scalebox{\tracescalex}[\tracescaley]{\includegraphics[page=129]{logos.pdf}}} & 
\end{tabular}
\label{tab:openpcrmotifs}
\end{table}    
}
\indent The context trace of these motifs hints towards two properties of the selection process. First, a clear selection towards single stranded regions for every extracted motif can be observed. It has always been stipulated that ssDNA/RNA binding motifs are most likely located in loop regions \cite{Schudoma2010}. Indeed, this assumption was leveraged by MEMERIS, by imposing structural priors directing the motif search towards single stranded regions. In the case of \aptastruct, no prior assumption of this type was made. The fact that despite a lack of such priors, motifs detected by \aptastruct\ conform with the expected properties of RNA sequence-structure binding sites support their relevance for binding. Next, the trend of the structural preferences of these motifs emerges relatively early during the selection process indicating that, in conjunction with our method, the identification of biologically relevant binding sites in general purpose HT-SELEX data might be possible with fewer selection cycles. 
\section{Conclusion}
Aptamers have a broad spectrum of applications and are increasingly being used to develop new therapeutics and diagnostics. HT-SELEX, in contrast to the traditional protocol, provides data for a global analysis of the selection properties and for simultaneous discovery of an large number of candidates. This extensive amount of information has utility only in conjunction with suitable computational methods to analyze the data.  \\
\indent Unlike in traditional SELEX, where only a handful of potential binders are retrieved and exhaustively tested experimentally, HT-SELEX returns a massive amount of sequencing data sampled from some, or all, selection rounds. This data consequently serves as the basis for the challenging task of identifying suitable binding candidates and for deriving their sequence-structure properties that are key for binding affinity and specificity. Except for the special case of TF binding aptamers, no previous tool addressing this task existed. The realization that a naive relationship between aptamer frequency and their binding affinity is not universally valid, further complicates this task. 
Several potential factors during any stage of the selection contribute to this 
complexity,
including polymerase amplification biases, sequencing biases, contamination of foreign sequences, and non-specific binding. These factors prompted aptamer experts to consider cycle-to-cycle enrichment instead of frequency counts as a predictor for binding affinity. While cycle-to-cycle enrichment did increase the predictive power of these methods, it cannot bypass problems related to amplification bias nor can it identify aptamer properties that drive binding affinity and specificity.\\
\indent In contrast, \aptastruct\ is specifically designed to identify sequence-structure binding motifs in HT-SELEX data and thus to predict the features behind binding affinity and specificity.  Importantly, rather than using quantitative information, it directly leverages the experimental design of the SELEX protocol and identifies motifs that are under selection through appropriately designed scoring functions. By focusing on local motifs that are selected for, \aptastruct \ bypasses global biases such as the PCR bias which is typically related to more universal sequence properties such as the CG content. In addition, because \aptastruct\ measures selection towards a given sequence-structure motif by the shift in the distribution of the structural context and not based on abundance, it can uncover statistically significant motifs that are selected for even when these only form a small fraction of the pool. This is an important property that can ultimately help to shorten the number of cycles needed for selection and thus to reduce the overall cost of the procedure. \\
\indent In testing on simulated data, \aptastruct\ outperformed other methods, in part because these methods models were not specifically designed to handle these sort of sequences. Furthermore, no competitors exist that could be tested on \textit{in vitro} data as none of the current programs scale to the amount of data points produced by HT-SELEX. While we currently have no gold standard for experimental data to measure the quality of the identified motifs, it is reassuring that the motifs converge structurally to loop regions, consistent with the accepted view where such binding sites reside. \\
\indent Sequence logos provide a convenient visualization of the selected motifs. For TF-binding, information used to derive these logos can be used to estimate binding energy \cite{stromo2002,beno2002}. However, for general HT-SELEX this connection is less immediate as one has to take into account the energy contribution from the structure component \cite{levens2015}. Perhaps even more importantly, aptamers binding to large cell surfaces are likely to be exposed to more binding opportunities than aptamers binding to single receptors and thus the number of resulting motifs can be expressed as a function of interaction probability and binding affinity. We hypothesize that the here presented $K$-context trace will be helpful in untangling some of these contributions.\\
\indent Finally, analysis of the $K$-context trace indicates that the selection signal can be identified at very early cycles. This  suggests that, with deep enough sequencing, only a limited number of selection cycles might be required. Yet, this analysis also shows that the dynamics of $K$-context traces is not the same for all sequence-structure motifs. While most trends essentially stabilize at a relatively early cycle, some continue to grow. We hypothesize that this type of information can aid the identification of the most promising binders.
Note that while we defined the $K$-context shifting score on all pairs of sequenced selection pools, it can also be used to focus the analysis to any part of the selection as long as it includes at least two cycles and hence to center on additional details of the selection dynamics. Other variants of the $K$-context shifting score (e.g. always using the initial pool as background/reference in the summation) can also  prove informative, however full elucidation of this dynamic will require concerted computational and experimental effort.
 \aptastruct\ is not only a powerful method to detect emerging sequence-structure  motifs but also, a flexible tool to interrogate such selection dynamics.

\newpage
\bibliographystyle{unsrt}

\newpage
\appendix
\begin{center}
\huge
{\bf Supplementary Material And Methods}
\end{center}
    
\section{SELEX Simulation Details} \label{sec:simulation}
In order to create data sets with properties comparable to \textit{in-vitro} HT-SELEX experiments, we designed a simulation scheme capable of mimicking target-specific selection, error-prone amplification, as well as the stochastic nature of partitioning each cycle into sequencing set and selection set. Furthermore, our simulation allows for the introduction of an arbitrary number of sequence-structure motifs of different sizes and affinities, and provides control over the binding strength of background sequences. The simulation builds on our AptaSIM software \cite{hoinkaNAR} and extends it to take secondary structure into account. For each aptamer $a$, we use  $|a|$, to denote its frequency and   $a_{bind}$ to denote its biding affinity.\\\\
For a user-defined pool size, a set of aptamers containing sequence-structure motifs of desired length is first produced using a second order hidden Markov model (HMM) trained with sequences from an \textit{in vitro} HT-SELEX experiment. To this end, we generate a sequence from the HMM, predict its secondary structure and identify all single-stranded regions larger or equal to the motif size. Next, we substitute the nucleotides in one of these unpaired regions with the motif and verify, using SFold, that the secondary structure profile of the motif region is predominantly single stranded. If the average probability of any of the single stranded secondary structure contexts in the motif region falls below 60\%, we discard this sequence, and otherwise add it to the pool. This procedure is repeated as many times as required to create the desired number of sequences containing the motifs. The remaining aptamers are sampled directly from the training data in order to ensure realistic secondary structure properties that sequences generated from the HMM might not necessarily possess.\\\\
Using this set of sequences as input, we perform a weighted sampling without replacement according to the aptamers count and affinity. I.e., until the desired sample size is reached, we compute a weight $w(a)$ for each aptamer $a$ as
\begin{equation}
	w(a) = \frac{|a|*a_{bind}}{\sum_{a \in R^x} |a|*a_{bind} },
\end{equation}
sample a sequence according to this distribution and adjust the weights to account for the removed aptamer. The resulting pool is then amplified by means of virtual PCR in which the amplification efficiency of the polymerase, as well as the mutation rate is adjustable. In order to simulate the sampling effect, we inject a user-defined percentage of singleton aptamers with low affinity, sampled from the training data, into the pool, and reduce the pool size to its original size by weighted sampling without replacement where the weights are generated according to the aptamer counts.\\\\
After each cycle, the pool is stored in fastq format and used as input for our \aptastruct\ algorithm.

\section{Generating Motif Logos and Choosing $k$} \label{sec:motifgeneration}
A $k$-mer does not necessarily cover the motif completely. In addition, aptamers can contain slightly modified sequences in the same structural context. Such motif variants can arise independently or as a result of polymerase errors during the amplification stages of the SELEX protocol. Therefor, our method clusters these elements in an iterative manner. First, $k$-mers with significant context shift scores are sorted in decreasing order according to their $k$-mer frequencies in the last selection cycle. We then select the top $k$-mer as a cluster seed, iterate through each of the remaining $k$-mers, and use the following decision scheme to choose whether or not to include the $k$-mer into to the cluster. Fully overlapping $k$-mers must have at most one mismatch with the seed. For partially overlapping $k$-mers and $k \leq 6$, the longest common substring (LCS) with the seed must be at least 4 nucleotides in length, while the LCS is required to be at least 5 for $k\geq 7$. Furthermore, their $K$-context trace must also satisfy similar structural shifts that correlate with the seed to be included in the cluster. To examine the similarities of the context shifts, for each $\mathbb{K}_i$, we identify the specific structural context $C'$ that changes the most over the selection rounds:
$$C'=\argmax_{C} f(C)=\sum_{x=1}^{m-1}{\sum_{y=x+1}^{m}{\mathbb{P}_i^y(C)-\mathbb{P}_i^x(C)}}$$ 
If $C'$ coincides with the dominant context of the seed, we include the $K$-context to the cluster.\\\\
The number of clusters that results from this approach can also serve as a measure for an appropriate choice of the $k$-mer size for a particular data set. By choosing $k$ such that it maximizes the number of the resulting clusters we ensure to capture all non-overlapping motifs in the pool. 

\section{CELL-SELEX Experimental Details} \label{sec:cellselex}
Cell-based SELEX was performed using an RNA library containing a randomized 30-nt region flanked by fixed primer sequences. Cell-based selection was performed as previously described [1], by employing open PCR for DNA amplification during each selection round. Positive selection was performed on Hela cells transduced with a bicistronic lentiviral vector expressing the target surface receptor and GFP, while unmodified Hela cells, which lack expression of the target receptor, were used for negative selection. High throughput sequencing (HTS) was performed on the positive selection at rounds 0, 1, 3, 5, 6, 7, 8, and 9. Significant molecular enrichment of receptor-specific aptamers was observed after five rounds of selection. 

\section{\aptastruct\ Results on \textit{in vitro} Data}
{
\newcommand{\motifscalex}{0.75}
\newcommand{\motifscaley}{0.30}
\newcommand{\tracescalex}{0.90}
\newcommand{\tracescaley}{0.30}
\newcommand{\rowspace}{4ex}
\rowcolors{2}{white}{blue!7}
\LTcapwidth=\textwidth
\begin{longtable}{m{3.5cm}m{2.5cm}m{3cm}m{2.2cm}m{4.1cm}m{0pt}}
\caption{Full set of the sequence-structure motifs as produced by \aptastruct\ on CELL-SELEX data. The sequence logo as well as the most frequent $k$-mer constituting the logo (Logo Seed) and its p-value are depicted for each motif. The $K$-context trace for the sequenced cycles (0,1,3,5,6,7,8,9) is shown in the last column.}
\endfirsthead
\endhead
\rowcolor{white} \bf Sequence Logo &\bf Logo Seed &\bf Seed p-Value &\bf Frequency &\bf Motif Context Trace \\[1pt] \hline
\rowcolor{white}& & & & \raisebox{-.5\height}{\scalebox{\tracescalex}[1]{\includegraphics{images/TRACERULER8.pdf}}} & \\
\raisebox{-.5\height}{\scalebox{\motifscalex}[\motifscaley]{\includegraphics[page=110]{logos.pdf}}}
& CTGTG & 5.31E-03 & 32.98 \% & 
\raisebox{-.5\height}{\scalebox{\tracescalex}[\tracescaley]{\includegraphics[page=109]{logos.pdf}}} & 
\\ [\rowspace]
\raisebox{-.5\height}{\scalebox{\motifscalex}[\motifscaley]{\includegraphics[page=124]{logos.pdf}}}
& TTATT & 7.16E-03 & 25.14 \% & 
\raisebox{-.5\height}{\scalebox{\tracescalex}[\tracescaley]{\includegraphics[page=123]{logos.pdf}}} & 
\\ [\rowspace]
\raisebox{-.5\height}{\scalebox{\motifscalex}[\motifscaley]{\includegraphics[page=126]{logos.pdf}}}
& GTTTA & 9.71E-03 & 16.77 \% & 
\raisebox{-.5\height}{\scalebox{\tracescalex}[\tracescaley]{\includegraphics[page=125]{logos.pdf}}} & 
\\ [\rowspace]
\raisebox{-.5\height}{\scalebox{\motifscalex}[\motifscaley]{\includegraphics[page=128]{logos.pdf}}}
& ATGTT & 3.66E-03 & 15.27 \% & 
\raisebox{-.5\height}{\scalebox{\tracescalex}[\tracescaley]{\includegraphics[page=127]{logos.pdf}}} & 
\\ [\rowspace]
\raisebox{-.5\height}{\scalebox{\motifscalex}[\motifscaley]{\includegraphics[page=130]{logos.pdf}}}
& GTGTC & 6.61E-03 & 13.95 \% & 
\raisebox{-.5\height}{\scalebox{\tracescalex}[\tracescaley]{\includegraphics[page=129]{logos.pdf}}} & 
\\ [\rowspace]
\raisebox{-.5\height}{\scalebox{\motifscalex}[\motifscaley]{\includegraphics[page=132]{logos.pdf}}}
& AATTG & 3.10E-03 & 12.76 \% & 
\raisebox{-.5\height}{\scalebox{\tracescalex}[\tracescaley]{\includegraphics[page=131]{logos.pdf}}} & 
\\ [\rowspace]
\raisebox{-.5\height}{\scalebox{\motifscalex}[\motifscaley]{\includegraphics[page=134]{logos.pdf}}}
& CTGGC & 2.46E-03 & 12.55 \% & 
\raisebox{-.5\height}{\scalebox{\tracescalex}[\tracescaley]{\includegraphics[page=133]{logos.pdf}}} & 
\\ [\rowspace]
\raisebox{-.5\height}{\scalebox{\motifscalex}[\motifscaley]{\includegraphics[page=136]{logos.pdf}}}
& CGCTG & 7.07E-03 & 11.30 \% & 
\raisebox{-.5\height}{\scalebox{\tracescalex}[\tracescaley]{\includegraphics[page=135]{logos.pdf}}} & 
\\ [\rowspace]
\raisebox{-.5\height}{\scalebox{\motifscalex}[\motifscaley]{\includegraphics[page=138]{logos.pdf}}}
& TAATG & 5.70E-04 & 10.12 \% & 
\raisebox{-.5\height}{\scalebox{\tracescalex}[\tracescaley]{\includegraphics[page=137]{logos.pdf}}} & 
\\ [\rowspace]
\raisebox{-.5\height}{\scalebox{\motifscalex}[\motifscaley]{\includegraphics[page=90]{logos.pdf}}}
& ATTAA & 2.64E-03 & 10.11 \% & 
\raisebox{-.5\height}{\scalebox{\tracescalex}[\tracescaley]{\includegraphics[page=89]{logos.pdf}}} & 
\\ [\rowspace]
\raisebox{-.5\height}{\scalebox{\motifscalex}[\motifscaley]{\includegraphics[page=92]{logos.pdf}}}
& TGCGC & 3.25E-04 & 9.73 \% & 
\raisebox{-.5\height}{\scalebox{\tracescalex}[\tracescaley]{\includegraphics[page=91]{logos.pdf}}} & 
\\ [\rowspace]
\raisebox{-.5\height}{\scalebox{\motifscalex}[\motifscaley]{\includegraphics[page=94]{logos.pdf}}}
& CTGCA & 1.49E-03 & 7.61 \% & 
\raisebox{-.5\height}{\scalebox{\tracescalex}[\tracescaley]{\includegraphics[page=93]{logos.pdf}}} & 
\\ [\rowspace]
\raisebox{-.5\height}{\scalebox{\motifscalex}[\motifscaley]{\includegraphics[page=96]{logos.pdf}}}
& AATAT & 1.90E-03 & 7.21 \% & 
\raisebox{-.5\height}{\scalebox{\tracescalex}[\tracescaley]{\includegraphics[page=95]{logos.pdf}}} & 
\\ [\rowspace]
\raisebox{-.5\height}{\scalebox{\motifscalex}[\motifscaley]{\includegraphics[page=98]{logos.pdf}}}
& CTTTG & 8.74E-03 & 6.59 \% & 
\raisebox{-.5\height}{\scalebox{\tracescalex}[\tracescaley]{\includegraphics[page=97]{logos.pdf}}} & 
\\ [\rowspace]
\raisebox{-.5\height}{\scalebox{\motifscalex}[\motifscaley]{\includegraphics[page=100]{logos.pdf}}}
& TGGTG & 4.65E-03 & 5.22 \% & 
\raisebox{-.5\height}{\scalebox{\tracescalex}[\tracescaley]{\includegraphics[page=99]{logos.pdf}}} & 
\\ [\rowspace]
\raisebox{-.5\height}{\scalebox{\motifscalex}[\motifscaley]{\includegraphics[page=102]{logos.pdf}}}
& GTAAA & 2.85E-03 & 4.79 \% & 
\raisebox{-.5\height}{\scalebox{\tracescalex}[\tracescaley]{\includegraphics[page=101]{logos.pdf}}} & 
\\ [\rowspace]
\raisebox{-.5\height}{\scalebox{\motifscalex}[\motifscaley]{\includegraphics[page=104]{logos.pdf}}}
& GCGTG & 3.87E-03 & 4.08 \% & 
\raisebox{-.5\height}{\scalebox{\tracescalex}[\tracescaley]{\includegraphics[page=103]{logos.pdf}}} & 
\\ [\rowspace]
\raisebox{-.5\height}{\scalebox{\motifscalex}[\motifscaley]{\includegraphics[page=106]{logos.pdf}}}
& TAAGT & 8.04E-03 & 3.95 \% & 
\raisebox{-.5\height}{\scalebox{\tracescalex}[\tracescaley]{\includegraphics[page=105]{logos.pdf}}} & 
\\ [\rowspace]
\raisebox{-.5\height}{\scalebox{\motifscalex}[\motifscaley]{\includegraphics[page=108]{logos.pdf}}}
& CATAA & 8.02E-03 & 3.74 \% & 
\raisebox{-.5\height}{\scalebox{\tracescalex}[\tracescaley]{\includegraphics[page=107]{logos.pdf}}} & 
\\ [\rowspace]
\raisebox{-.5\height}{\scalebox{\motifscalex}[\motifscaley]{\includegraphics[page=112]{logos.pdf}}}
& GTTAG & 8.60E-03 & 3.70 \% & 
\raisebox{-.5\height}{\scalebox{\tracescalex}[\tracescaley]{\includegraphics[page=111]{logos.pdf}}} & 
\\ [\rowspace]
\raisebox{-.5\height}{\scalebox{\motifscalex}[\motifscaley]{\includegraphics[page=114]{logos.pdf}}}
& ATAGT & 6.15E-03 & 2.91 \% & 
\raisebox{-.5\height}{\scalebox{\tracescalex}[\tracescaley]{\includegraphics[page=113]{logos.pdf}}} & 
\\ [\rowspace]
\raisebox{-.5\height}{\scalebox{\motifscalex}[\motifscaley]{\includegraphics[page=116]{logos.pdf}}}
& TGCCG & 2.78E-04 & 2.61 \% & 
\raisebox{-.5\height}{\scalebox{\tracescalex}[\tracescaley]{\includegraphics[page=115]{logos.pdf}}} & 
\\ [\rowspace]
\raisebox{-.5\height}{\scalebox{\motifscalex}[\motifscaley]{\includegraphics[page=118]{logos.pdf}}}
& TGCTC & 6.10E-03 & 2.52 \% & 
\raisebox{-.5\height}{\scalebox{\tracescalex}[\tracescaley]{\includegraphics[page=117]{logos.pdf}}} & 
\\ [\rowspace]
\raisebox{-.5\height}{\scalebox{\motifscalex}[\motifscaley]{\includegraphics[page=120]{logos.pdf}}}
& GTATG & 9.05E-03 & 2.30 \% & 
\raisebox{-.5\height}{\scalebox{\tracescalex}[\tracescaley]{\includegraphics[page=119]{logos.pdf}}} & 
\\ [\rowspace]
\raisebox{-.5\height}{\scalebox{\motifscalex}[\motifscaley]{\includegraphics[page=122]{logos.pdf}}}
& CCTAT & 2.27E-03 & 1.53 \% & 
\raisebox{-.5\height}{\scalebox{\tracescalex}[\tracescaley]{\includegraphics[page=121]{logos.pdf}}} & 
\label{tab:fullopenpcrmotifs}
\end{longtable}    
}

\section{Parameters used in this Study} \label{sec:parameters}
\textbf{\aptastruct:} We define all aptamers as low count that have a frequency lower than 3, i.e. we define $\alpha = 3$.\\\\
\textbf{Simulation:} We created a simulated dataset of 4 million sequences over 10 selection cycles. The initial round was designed to contain 5 motifs between 5-8 nucleotides in length and with binding affinities of 50\%, 55\%, 60\%, 70\%, and 75\% respectively (the background affinities were randomly assigned in a range of 1-25\%). Each motif was initially represented a total of 100 times in the pool. In each cycle, we sample a total of 40\% of the pool, amplify the selected aptamers with an efficiency of 95\% and a mutation rate of 5\%, and introduce a total of 50\% background sequences into the mix.\\\\
\textbf{Comparison with other Methods:} The $k$-mer size for DREME was restricted to the range $k=5,\dots,8$ and we limited it to output the top 100 PWM models with an e-value of at most $0.01$. For RNAcontext, we utilized the log of the frequency of each sequence to represent their binding affinities. Finally, we report the consensus sequence for each output PWM from DREME, RNAcontext and AptaTRACE. To compute the value for $C^*$, we consider a PWM as a matching an implanted motif if the sequence of the implanted motif is a substring of the consensus sequence of the PWM.

\section{Implementation Details and Runtime}
Our \aptastruct\ pipeline is implemented as a modular system in C++ and Java. RNA secondary structure profiles for all aptamers in the pool were predicted in parallel on the server farm at the National Center for Biotechnology Information, NIH, using 1000 Cores and 200MB or RAM per job. The simulated data required about 5 hours of wall clock time, whereas the \textit{in vitro} selection data finished in roughly 4 days. The enumeration of $k$-mers and consequent extraction of $k$-contexts was implemented as a multi-threaded C++ library and is capable of processing the here presented results in approximately 10 hours for all choices of $k$ between 5-8 on a large memory server with 100 CPUs. Finally, the context shifting scores, $k$-mer extraction, and clustering is implemented as a multi-threaded JAVA program requiring 3-4 hours, depending on the data size, for its completion.\\\\
The source code, is currently available on request and will be hosted online at:\\ \url{www.ncbi.nlm.nih.gov/CBBresearch/Przytycka/index.cgi\#aptatools}
\end{document}